\documentclass{PoS}

\usepackage{graphicx}
\usepackage{subfigure}

\newcommand{\be}{\begin{equation}}
\newcommand{\ee}{\end{equation}}

\newcommand{\bea}{\begin{eqnarray}}
\newcommand{\eea}{\end{eqnarray}}
\newcommand{\dis}{\displaystyle}

\title{Equation of State at Finite Density from Imaginary Chemical Potential}

\ShortTitle{Equation of State at Finite Density from Imaginary Chemical Potential}

\author{\speaker{Tetsuya Takaishi}\\
        Hiroshima University of Economics\\
        Hiroshima 731-0192  JAPAN\\
        E-mail: \email{takaishi@hiroshima-u.ac.jp}}

\author{Philippe de Forcrand\\
        Institut f\"ur Theoretische Physik, ETH Z\"urich, CH-8093 Z\"urich, SWITZERLAND\\
and \\
        CERN, Physics Department, TH Unit, CH-1211 Geneva 23,  SWITZERLAND \\
        E-mail: \email{forcrand@phys.ethz.ch}}

\author{Atsushi Nakamura\\
        Hiroshima University\\
        Higashi-Hiroshima 739-8521 JAPAN\\
        E-mail: \email{nakamura@riise.hiroshima-u.ac.jp}}

\abstract{
We perform two flavor QCD simulations with an imaginary chemical potential
and measure derivatives of the pressure up to 4th order as a function
of the imaginary chemical potential and the temperature 
$T \in [0.83 T_c, 2 T_c]$.
For temperatures $T \geq T_c$, these derivatives are fitted 
by a Taylor series in $\mu/T$ about $\mu=0$. 
A fit limited to 4th order describes the data poorly 
at all temperatures, showing that we are sensitive to 6th order contributions.
Similarly, a 6th order fit fails for temperatures $T_c \leq T \leq 1.05 T_c$, showing
the need for 8th order terms. Thus, our method may offer a computational
advantage over the direct measurement of Taylor coefficients at $\mu=0$.
At temperatures $T \leq T_c$, we fit our data with a hadron resonance gas 
ansatz. The fit starts to fail at $T \gtrsim 0.95 T_c$.
Using our fits, we also reconstruct the  equation of state as a function of 
real quark and isospin chemical potentials.
          }

\FullConference{The XXVII International Symposium on Lattice Field Theory - LAT2009\\
		 July 26-31 2009\\
		 Peking University, Beijing, China}

\begin{document}

\section{Introduction}
Although lattice QCD has been used successfully for simulations at zero and finite temperatures
and at zero density, 
Monte Carlo simulations at non-zero densities suffer from 
a technical problem: the lattice QCD action becomes complex, which
prevents its customary probabilistic interpretation. 
In principle one could perform simulations at zero density, and use the 
reweighting technique to obtain information at finite densities.
An early attempt known as the Glasgow method~\cite{Glasgow} did not work
due to the overlap problem: the configurations at zero density were too
``far'' from the target configurations at non-zero densities. 
Considerable progress has been accomplished by generalizing the Glasgow method 
to two-parameter reweighting~\cite{Fodor}.
Nevertheless, the range of reliability of this technique is difficult to 
assess, and its failure can go undetected.

Therefore, another, more conservative way to deal with  finite baryon densities 
may be useful. It consists of calculating Taylor coefficients of observables 
with respect to the chemical potential $\mu$ about $\mu=0$.
Those Taylor coefficients can be expressed as expectation values of 
complicated observables, which can be measured at zero density. Thus, 
there is no difficulty to perform Monte Carlo simulations in this method.
A first, pioneering attempt to obtain quark susceptibilities~\cite{Gottlieb} 
has been followed by numerous works, obtaining in particular 
the response of screening masses to chemical potential~\cite{QCDTARO,QCDTARO2,QCDTARO3}.
The Taylor expansion method has also been used for 
studies of the equation of state, of the phase transition and of higher order susceptibilities~\cite{Allton,Allton2,Gupta,Gupta2}.
However, the complexity of the observable representing the Taylor coefficient, and the computer
effort to measure it, increase rapidly with the order of the Taylor expansion.
This motivates us to follow a different strategy.

Since no difficulty appears for simulations at imaginary chemical potential $\mu=i \mu_I$,
one can obtain information at finite baryon densities by analytic continuation of
observables measured at finite $\mu_I$.
Actually, this imaginary chemical potential strategy has been applied 
with success to the determination of the phase transition~\cite{Forcrand}.  

In this study, we perform simulations at finite $\mu_I$  and
measure derivatives of the pressure {\em as a function of $\mu_I$}.
These derivatives contain information about the Taylor coefficients
of the $\mu=0$ expansion, which can be extracted by fitting.
Finally, we try to reconstruct the equation of state at finite baryon and isospin densities.
The strategy of our method and preliminary results were presented rather long
ago in \cite{GENOA}.
Here we report further progress on this project. 
A related approach, where the quark density is measured at imaginary quark and real isospin chemical potentials and then fitted by a polynomial ansatz, has recently been presented in
\cite{DELIA}.

\section{Equation of State at Finite Chemical Potential}

The lattice QCD partition function with $N_f$ flavors of staggered fermions can be written as
\be
Z=\int \Pi_i^{N_f} \det M(U,m_i,\mu_i)^{1/4}\exp(-S_g[U])dU,
\ee
where $S_g[U]$ is the gauge action and $M(U,m_i,\mu_i)$ stands for 
the staggered Dirac operator with quark mass $m_i$ and chemical potential $\mu_i$\footnote{We set aside potential problems with ``rooting'' the determinant,
particularly at non-zero chemical potential.}.
In this study we consider $N_f=2$ degenerate fermion species
and use the standard Wilson gauge action.  

The pressure or the equation of state with chemical potential $\mu_u$ and $\mu_d$ is given by
\be
p(\mu_u,\mu_d) =-\frac{F}{V}=\frac{T}{V}\ln Z(\mu_u,\mu_d),
\ee
and can be expanded in a Taylor series about $\mu_u=\mu_d=0$ as  
\be
\frac{\Delta p}{T^4} \equiv \frac{p(\mu_u,\mu_d)- p(0,0)}{T^4} = 
\sum_{n,m=1}\frac1{n!m!}f_{nm}\left(\frac{\mu_u}{T}\right)^n \left(\frac{\mu_d}{T}\right)^m,
\ee
where $f_{nm}$ are the Taylor expansion coefficients.
They vanish when $(n+m)$ is odd due to CP symmetry.
Furthermore, for equal quark masses there is another symmetry $f_{nm}=f_{mn}$.
The $f_{nm}$'s are related to derivatives $\chi_{ij}$ of the pressure measured
at {\em non-zero} chemical potential by 
\be
T^{i+j-4} \chi_{ij}=\frac{\partial^{i+j} (p(\mu_u,\mu_d)/T^4)}
{\partial(\mu_u/T)^i \partial(\mu_d/T)^j}
=\sum_{n=i,m=j}\frac1{(n-i)!(m-j)!}f_{nm}\left(\frac{\mu_u}{T}\right)^{n-i} \left(\frac{\mu_d}{T}\right)^{m-j}.
\label{Deri}
\ee
While at zero density $\chi_{ij}=f_{ij} T^{4-i-j}$, 
at non-zero densities $\chi_{ij}$ includes higher order $f_{nm}$ terms,
and does not vanish for odd $(i+j)$.
This suggests to 
use all available $\chi_{ij}$'s at non-zero densities, in order to estimate the $f_{nm}$'s.
Here, we try to estimate ${f_{nm}}$ by fitting all $\chi_{ij}$ simultaneously to the polynomial expansions eq.(\ref{Deri}). 
Of course, $\chi_{ij}$ at non-zero baryon density is not directly obtainable from simulations on the lattice because of the sign problem.
However, $\chi_{ij}$ can be obtained through simulations at imaginary quark chemical potential or at real isospin density.
Here, we calculate $\chi_{ij}$ at imaginary chemical potentials.

%
%

Therefore, we set $\mu=i\mu_I$.
Each $\chi_{ij}$ depends on higher order Taylor coefficients following eq.(\ref{Deri}).
Therefore, with sufficiently accurate data on $\chi_{ij}$ one can also obtain higher order Taylor coefficients $f_{nm}$, $n>i, m>j$.
The measurements of the derivatives involve computing traces of inverse Dirac matrix products.
These traces were estimated using the noise method with 40 $Z_2$ random vectors.
In this study we measure $\chi_{ij}$ up to $i+j=4$. Thus we have 8 different $\chi_{ij}$'s.
We fit all the data to the corresponding 8 polynomial expansions eq.(\ref{Deri}) truncated to
a given order $(n+m)$, and try to obtain
the Taylor coefficients $f_{nm}$.

As we will see, in the confined phase a Taylor expansion is not the most compact description 
of the pressure.
Instead, for $T\leq T_c$ we use the Hadron Resonance Gas (HRG) model.
In the HRG model the pressure is given as\footnote{This expression is taken from (4.3) in \cite{Allton2}.}
\bea
\label{HRG}
\frac{\Delta p(\mu_u,\mu_d)}{T^4} & = &G[\cosh(\frac{2\mu_{Is}}{T})-1] +R[\cosh(\frac{3\mu_q}{T})\cosh(\frac{\mu_{Is}}{T})-1] \\ 
&+&W[\cosh(\frac{3\mu_q}{T})\left(\cosh(\frac{\mu_{Is}}{T})+\cosh(\frac{3\mu_{Is}}{T})\right)-2], \nonumber
\eea
where $G$,$R$ and $W$ are constants related to the hadron spectrum, 
and 
quark and isospin chemical potentials $\mu_q$ and $\mu_{Is}$ are defined as
$\mu_q=(\mu_u+\mu_d)/2$ and $\mu_{Is}=(\mu_u-\mu_d)/2$ respectively.

The derivatives of the pressure with respect to $\mu_u$ and $\mu_d$, 
instead of having the polynomial form eq.(\ref{Deri}), are now obtained by differentiating eq.(\ref{HRG}).
The coefficients $G$,$R$ and $W$ are then extracted by fitting imaginary-$\mu$
data.
In terms of $G$,$R$ and $W$, the first Taylor coefficients are given by
\be
f_{20}=\left(G+\frac52 R +7W\right),
\ee
\be
f_{11}=-\left(G+2(R+W)\right),
\ee
\be
f_{22}=G+4(R+W),
\ee
\be
f_{31}=G+\frac72 R+ 49W,
\ee
\be
f_{40}=-G+5(R+W).
\ee

\begin{table}[h]
  \centering
  \caption{$\chi^2/dof$ for polynomial ansatz of degree 4, 6 and 8 (maximum value of $(n+m)$ in eq.(\protect\ref{Deri})).}
  \label{tab:1}
   \begin{tabular}{cllllllllllllllll}
 \hline
$T/T_c$   & 0.99 & 1.00 & 1.03 & 1.04 & 1.065 & 1.085  & 1.1 & 1.2 & 1.3 & 1.4 & 1.5 & 2.0 \\ 
 \hline
4th       & 85.1 & 134.9& 3.15 & 3.14 & 3.50 & 7.24 & 3.20 & 10.9 & 11.3 & 6.67 & 9.95 & 9.15 \\
6th       & 19.1 & 42.1 & 1.60 & 2.19 & 0.82 & 5.50 & 5.53 & 1.52 & 0.89 & 2.46 & 1.09 & 2.10 \\
8th       & 4.53 & 5.29 & 1.64 & 1.77 & 0.81 & 1.01 & 2.15 & 1.72 & 0.91 & 2.16 & 1.22 & 1.28 \\
\hline
    \end{tabular}
\end{table}

\begin{table}[h]
  \centering
  \caption{$\chi^2/dof$ for HRG ansatz.} 
  \label{tab:2}
   \begin{tabular}{clllllllllllllll}
 \hline
$T/T_c$    & 0.83 & 0.9 & 0.95 & 0.98 & 0.99 & 1.00   \\
 \hline
HRG       & 1.29 & 1.00 & 2.10 & 15.8 & 10.5 & 29.4    \\
\hline
    \end{tabular}
\end{table}


\begin{figure}
\vspace{5mm}
\centering
\includegraphics[height=5.5cm]{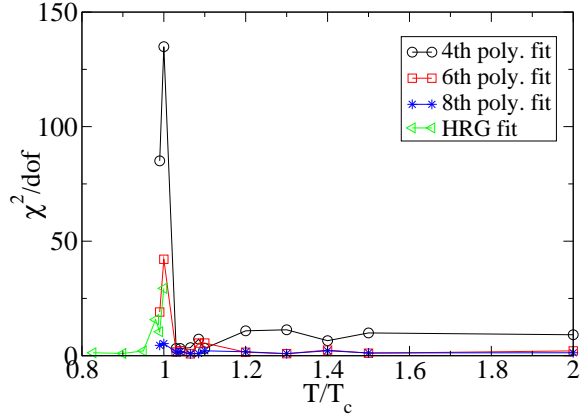}
\caption{
$\chi^2/dof$ of various polynomial ans\"atze as a function of $T/T_c$. 
The fitting range of $a\mu_I$ is 0.0-0.24.}
\label{fig:chisq2}
\end{figure}

\begin{figure}
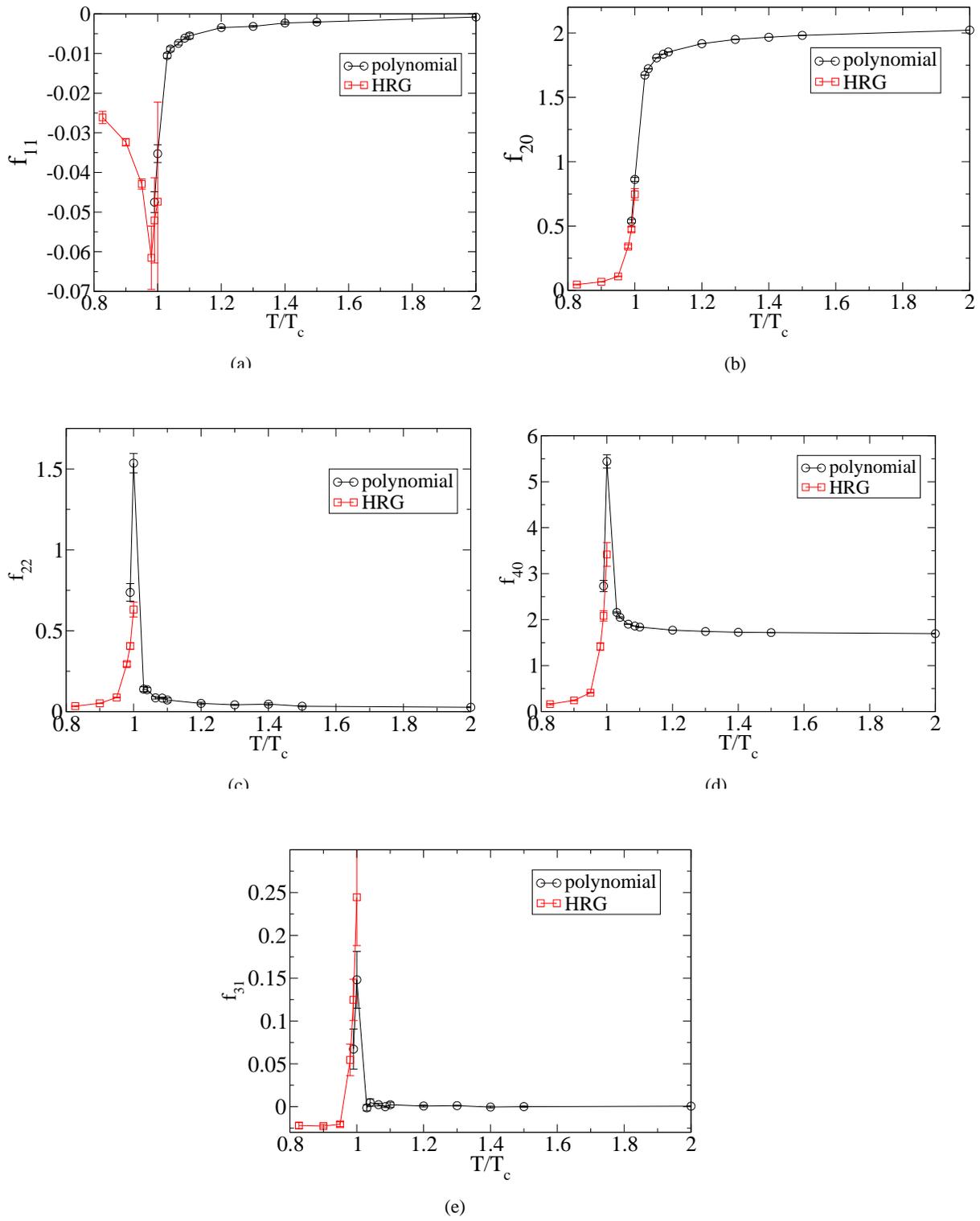

\vspace{5mm}
\centering
\subfigure[]{
\includegraphics[height=5.5cm]{c11all2-2.eps}
}~
\subfigure[]{
\includegraphics[height=5.5cm]{c20all2-2.eps}
} \\
\vspace{5mm}
\subfigure[]{
\includegraphics[height=5.5cm]{c22all2.eps}
}~
\subfigure[]{
\includegraphics[height=5.5cm]{c40all2.eps}
} \\
\vspace{5mm}
\subfigure[]{
\includegraphics[height=5.5cm]{c31all2.eps}
}
\caption{
Taylor coefficients: (a)$f_{11}$, (b)$f_{20}$, (c)$f_{22}$, (d)$f_{40}$ and (e)$f_{31}$.}
\label{fig:Taylor}
\end{figure}

\begin{figure}
\vspace{-5mm}
\centering
\subfigure[]{
\includegraphics[height=5.5cm]{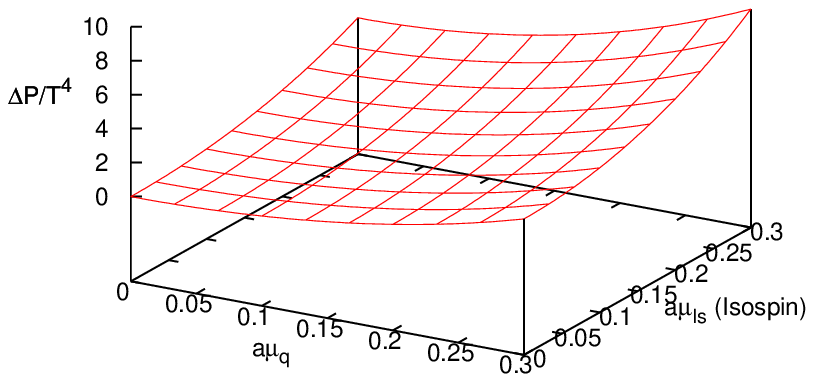}
}~
\subfigure[]{
\includegraphics[height=5.5cm]{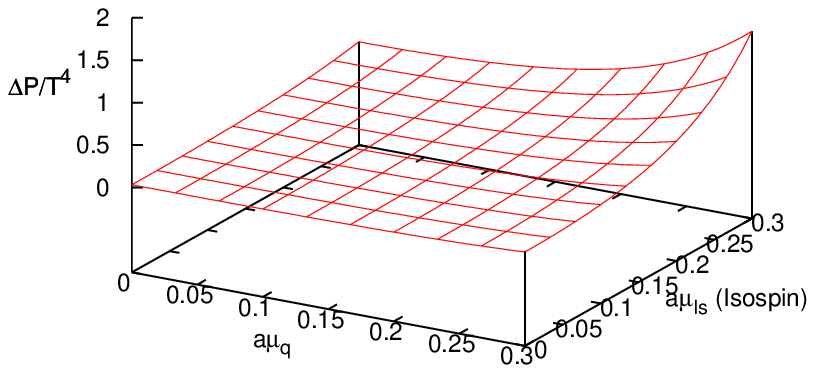}
}
\caption{
Equation of state (Pressure) as a function of $a\mu_q$ and $a\mu_{Is}$ at (a) $\beta=5.53 (T/T_c\sim 1.1)$ and at (b)$\beta=4.90 (T/T_c\sim 0.83)$.}
\label{fig:Pb553}
\end{figure}

%

\section{Simulations at Imaginary Chemical Potential}
We have performed simulations on $8^3 \times 4$ lattices
at a quark mass $m_q=0.05$ and imaginary chemical potentials $a\mu_I=0.0,\dots,0.24$.
We have chosen 16 values of $\beta$ ranging from 4.90 to 6.85, which correspond to
$T/T_c=0.83 \sim 2.0$.
Most of the simulations were performed using
the R-algorithm with a step size $\Delta t=0.02$. We also used the Rational HMC algorithm~\cite{RHMC} to
check the systematic stepsize errors caused by the R-algorithm, and found  
no significant difference for this lattice size and quark mass. 
At each simulation point 
we have accumulated 12000 to 20000 measurements. 
The measurements were taken every 5 trajectories
to balance the computational effort of the R-algorithm simultion and measurements.

\subsection{Fitting to $\chi_{ij}$}
We determine $f_{nm}$ by fitting all the derivatives
simultaneously to the corresponding ansatz of $\chi_{ij}$.
We used the polynomial ansatz eq.(\ref{Deri}) for the data at $T/T_c\geq0.99$, 
and the HRG ansatz  eq.(\ref{HRG}) at $T/T_c\leq1.0$.
Tables 1 and 2 show the $\chi^2/dof$ for the polynomial and HRG fits, respectively.
The fitting range of $a\mu_I$ is $0.0-0.24$, which covers most of the range up to
the Roberge-Weiss transition at $\mu_I = \pi T/3$.

Fig.\ref{fig:chisq2} compares the $\chi^2/dof$ among various polynomial and HRG fits.
One can see that the 4th order polynomial ($(n+m)\leq 4$ in the expansion eq.(\ref{Deri}))
is not good over the whole temperature range, and
that the 6th order one becomes poor in the vicinity of $T_c$. 
Similarly, one can also see that the quality of the fit based on the HRG ansatz becomes poor 
for $T/T_c\ge 0.95$.
While the failure of the HRG ansatz near $T_c$ has been noticed before~\cite{Cheng,DELIA},
it is remarkable that we can see clear indications of 6th order, and even 8th order 
Taylor coefficients with our modest study. The measurement of 8th order Taylor coefficients
represents the current state of the art~\cite{Gupta2}. 

Fig.\ref{fig:Taylor} shows the Taylor coefficients $f_{11}$, $f_{20}$, $f_{22}$, $f_{40}$ and $f_{31}$ as a function of temperature.
Those results are obtained by fitting a 6th order polynomial in a range of $a\mu_I=0.0-0.24$. 
For $T/T_c\ge 1.0$, they agree well with those obtained  
from the direct measurement of derivatives at $\mu=0$, i.e. $\chi_{ij}|_{\mu=0}$, 
but are more accurate.
We do not show the 6th order Taylor coefficients $f_{60}, f_{51}, f_{42}$ and $f_{33}$: even though their collective effect is statistically significant, they cannot be individually 
determined with any statistical accuracy. We only observe that $f_{60}$ is dominant at this
order.

Similary, the Taylor coefficients obtained from the HRG ansatz for $T/T_c\le 0.95$
also agree well with direct measurements of $\chi_{ij}|_{\mu=0}$, with higher accuracy.
However, for $T/T_c > 0.95$  the results from the HRG ansatz fits deviate from  $\chi_{ij}|_{\mu=0}$.
This observation is consistent with the measured $\chi^2/dof$, which increase considerably for $T/T_c > 0.95$.

\subsection{Equation of State at Finite Densities} 
Once we obtain the Taylor coefficients of the pressure or the parameters of the HRG model,
we can reconstruct the equation of state.
Here, we present two cases at $\beta=5.53(T/T_c\sim 1.1)$ and $\beta=4.90(T/T_c\sim 0.83)$
which are reconstructed with the Taylor series and the HRG ansatz, respectively.
Fig.~\ref{fig:Pb553}(a) shows the equation of state
at $\beta=5.53 (T/T_c\sim 1.1)$ as a function of $a\mu_q$ and $a\mu_{Is}$.
Similarly Fig.~\ref{fig:Pb553}(b) shows the equation of state
at $\beta=4.90 (T/T_c\sim 0.83)$.

\begin{figure}
\vspace{-5mm}
\centering
\subfigure[]{
\includegraphics[height=5.5cm]{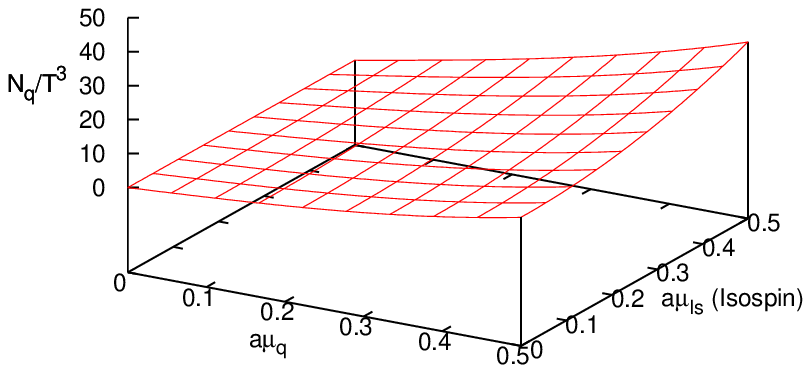}
}~
\subfigure[]{
\includegraphics[height=5.5cm]{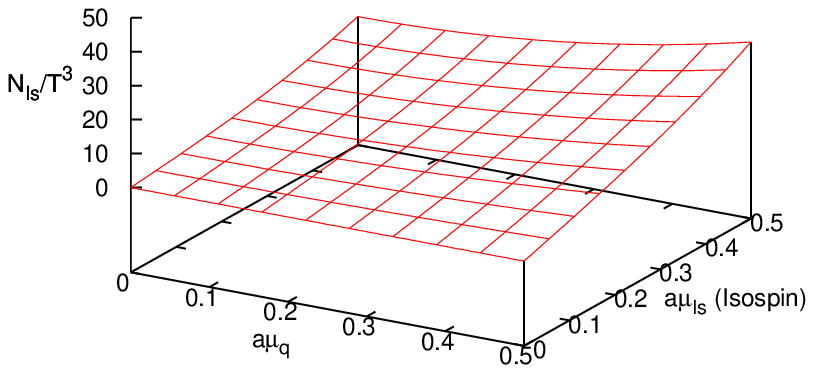}
}
\caption{
Number density at $\beta=5.53 (T/T_c\sim 1.1)$ as a function of $a\mu_q$ and $a\mu_{Is}$: (a)$N_q/T^3$ and (b) $N_{Is}/T^
3$.}
\label{fig:Numb553}
\end{figure}

\begin{figure}
\vspace{-5mm}
\centering
\subfigure[]{
\includegraphics[height=5.5cm]{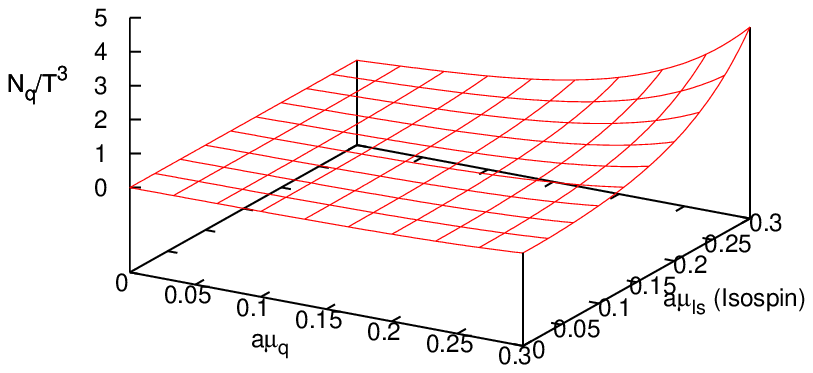}
}~
\subfigure[]{
\includegraphics[height=5.5cm]{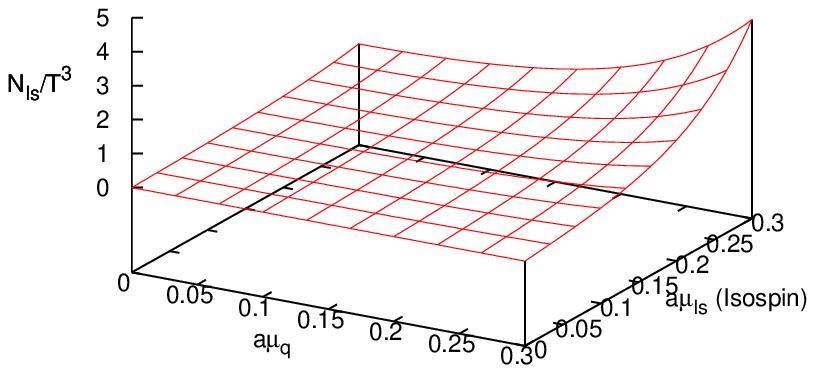}
}
\caption{
Number density at $\beta=4.90 (T/T_c\sim 0.83)$ as a function of $a\mu_q$ and $a\mu_{Is}$: (a)$N_q/T^3$ and (b) $N_{Is}/T
^3$.}
\label{fig:Numb490}
\end{figure}

One can also reconstruct other interesting quantities. 
Fig.~\ref{fig:Numb553} shows the quark number density $N_q$ and the isospin number density $N_{Is}$
at $\beta=5.53$ as a function of $a\mu_q$ and $a\mu_{Is}$.  
Similarly, Fig.~\ref{fig:Numb490} shows  $N_q/T^3$ and $N_{Is}/T^3$ at $\beta=4.90$.
Here,
$N_q$ and $N_{Is}$ are defined as
$\dis N_q = \frac{\partial p}{\partial \mu_q}$ and
$\dis N_{Is}= \frac{\partial p}{\partial \mu_{Is}}$, respectively.

\section{Conclusions}
We have performed simulations at imaginary chemical potentials and
measured the derivatives of the pressure with respect to $\mu$, at zero and non-zero imaginary $\mu$.
By fitting all the derivatives to a polynomial ansatz or an HRG ansatz,
we obtained the Taylor coefficients of the $\mu/T$ expansion of the pressure about $\mu=0$.
The Taylor coefficients obtained by a polynomial fit for $T/T_c \ge 1.0$ 
agree well with the direct measurement of derivatives at $\mu=0$, $\chi_{ij}|_{\mu=0}$,
but are more accurate. 
Remarkably, we find it impossible to obtain a good fit, at any temperature, without including
6th order derivatives. For $T_c \leq T \leq 1.04 T_c$, 8th order derivatives are necessary.
Thus, our approach may provide a cheaper alternative to the direct measurement of high-order
derivatives at $\mu=0$.

Similarly, below $T_c$ we observed that the Taylor coefficients obtained by the HRG ansatz 
deviate from $\chi_{ij}|_{\mu=0}$ for $T/T_c \ge 0.95$, and the HRG ansatz itself gives a 
poor description of the imaginary-$\mu$ data.
The same observation has been made in \cite{DELIA}.

Finally, using the obtained Taylor coefficients we reconstructed the equation of state
and the number densities as a function of $\mu_q$ and $\mu_{Is}$ up to 4th order.

\section*{Acknowledgements}
The numerical calculations were carried out on SX8 at YITP in Kyoto University.


\end{document}